# Quantum Size Effect in Optically Active Indium Selenide Crystal Phase Heterostructures Grown by Molecular Beam Epitaxy

*Piotr Wojnar\*, Maciej Wójcik, Piotr Baranowski, Jacek Kossut, Marta Aleszkiewicz, Jarosław Z. Domagała, Róża Dziewiątkowska, Paweł Ciepielewski, Maksymilian Kuna, Zuzanna Kostera, Sławomir Kret, Sergij Chusnutdinow*

P. Wojnar, M. Wójcik, J. Kossut, M Aleszkiewicz, J.Z. Domagała, R. Dziewiątkowska, P. Ciepielewski, S. Kret, S. Chusnutdinow
*Institute of Physics, Polish Academy of Sciences,*
*Al Lotników 32/46, PL 02-668 Warsaw, Poland*
Emails: wojnar@ifpan.edu.pl, mwojcik@ifpan.edu.pl, kossut@ifpan.edu.pl, mwesol@ifpan.edu.pl, domag@ifpan.edu.pl, rozad@magtop.ifpan.edu.pl, pawciep@gmail.com, kret@ifpan.edu.pl, xystik@gmail.com

P. Wojnar, R. Dziewiątkowska, S. Chusnutdinow
*International Research Centre MagTop,*
*Institute of Physics, Polish Academy of Sciences,*
*Al Lotników 32/46, PL 02-668 Warsaw, Poland*

P. Baranowski
*Institute of Applied Physics, Military University of Technology,*
*ul. gen. Sylwestra Kaliskiego 2, PL 00 – 908 Warsaw, Poland*
piotr.baranowski@wat.edu.pl

P. Ciepielewski
*Łukasiewicz Res Network, Institute of Microelectronics & Photonics,*
*Al Lotników 32/46, PL 02-668 Warsaw, Poland*

*VIGO Photonics S.A.,*
*ul. Poznańska 129/133, 05-850 Ożarów Mazowiecki, Poland\*

M. Kuna
*Institute of Experimental Physics, Faculty of Physics, University of Warsaw,*

ul. Pasteura 5, PL 02-093 Warsaw, Poland

mc.kuna@student.uw.edu.pl

Z. Kostera

*Faculty of Physics, Warsaw University of Technology,*

*Koszykowa 75, PL 00-662 Warsaw, Poland*

zuzanna.kostera.stud@pw.edu.pl

**Abstract**

Indium selenide attracts the interest due to its outstanding electronic and optical properties which are potentially prospective in view of applications in electronic and photonic devices. Most of the polymorphic crystal phases of this semiconductor belong to the family of two-dimensional van der Waals semiconductors. In this study optically active indium selenide crystal phase heterostructures are fabricated by molecular beam epitaxy in a well-controlled manner. It is demonstrated that by changing the growth conditions one may obtain either γ-InSe, or γ-$In_2Se_3$, or β-y$In_2Se_3$ crystal phases. The most promising crystal phase heterostructures from the point of view of photonic applications is found to be the γ-InSe/γ-$In_2Se_3$ heterostructure. An intense optical emission from this heterostructure appears in the near infrared spectral range. The emission energy can be tuned over 250 meV by changing γ-InSe layer thickness which is explained by the quantum size effect. The optically active indium selenide crystal phase heterostructures represent, therefore, an interesting platform for the design of light sources and detectors in the near infra-red. The use of molecular beam epitaxy for this purpose ensures that the structures are fabricated on large surfaces opening the possibility for the design of device prototypes by using lithography methods.

# 1. Introduction

The crystalline structure of two-dimensional (2D) semiconductors is characterized strong in-plane covalent bonds and weak interlayer van der Waals forces. These materials are considered as building blocks of next generation electronics and photonics. The ability of defect-free stacking of different materials thin layers on top of each other and the effective tuning of electronic and photonic properties by changing the number of layers, their chemical composition and the stacking order ensures that they represent a versatile platform for innovative quantum science technology[1,2]. The exfoliation technique used for the fabrication of the most reported van der Waals heterostructures results in structures with not well-controlled in-plane shapes. On the other hand, in the case of the van der Waals heterostructures grown by epitaxial methods, e.g., molecular beam epitaxy (MBE) used in this work, the structures are fabricated on large areas with clean interlayer interfaces and good repeatability of their physical properties opening the path for their future industrial applications.

Indium selenide attracts a great interest due to its outstanding electronic and photonic properties, such as the excellent photo-responsivity[3–5], high electron mobility[6–9] robust room temperature ferroelectricity[10], high Seebeck coefficient and the largest band gap tunability among 2D materials depending on the number of layers[11]. One of the challenges of fabrication indium selenide on large scale substrates is the complexity of indium selenide crystal phase diagram[12–15]. Different crystalline phases of indium selenide are characterized by significantly different values of the fundamental band gap[10,16–19], for instance, the room temperature band gap of bulk $\gamma$-$In_2Se_3$ amounts to 1.95 eV and that of bulk $\gamma$-InSe to 1.26 eV. Therefore, it is in principle possible to tune the optical emission from heterostructures built from different indium selenide crystal phases in a broad spectral range: from the near infra-red to the visible range depending on the number of layers and the crystal phase. While MBE has been employed to fabricate indium selenide layers with various crystalline structures[10,20–22], the former papers are related mostly to the properties of bulk-like $\gamma$-$In_2Se_3$ and $\alpha$-$In_2Se_3$ phases[20–22]. More recent publications report on the growth of thin films of $\alpha$-$In_2Se_3$ for the study of their ferroelectric properties[10], and on the deposition of a few layers $\gamma$-InSe for the observation of 2D electron gas by scanning tunneling microscopy[7].

In this work, we exploit the opportunity to fabricate heterostructures built entirely of different indium selenide crystal phases in an MBE process by changing the growth conditions in a controlled manner. The special focus is put on the optical properties of these structures. Combinations of indium selenide phases characterized by intensive optical transitions in the near infra-red are identified. It is demonstrated that the emission energy can be effectively tuned over a broad spectral range by changing the parameters of heterostructures, such as the layer thicknesses, which is related to the quantum size effect.

## 2. Results and Discussion

### 2.1 Crystal phase control of indium selenide

In the first step, MBE technique is used to gain the control over the indium selenide crystal phase. The setup is equipped with separate indium and selenium flux sources. The indium flux is chosen to be in the range $1 - 4 \cdot 10^{-8}$ torr and the selenium flux varies from $1 \cdot 10^{-8} - 3 \cdot 10^{-6}$ torr depending on the sample. The substrate of choice is (111)B-oriented GaAs because its interatomic distance in the (111)-plane, 3.997 Å, is similar to the lattice constants of the most of the indium selenide crystalline phases. After thermal deoxidation of the substrate at 590 ºC (without any fluxes from the sources) the growth temperature is stabilized in the range 300-450 ºC, depending on the process. For the indium selenide growth indium and selenium fluxes are opened simultaneously. In the MBE process there are generally two main growth parameters: the growth temperature and the In/Se flux ratio. The investigation of about 40 samples grown at different growth conditions resulted in the indium selenide phase diagram presented in Figure 1a. Importantly, the optimal growth conditions for three different indium selenide crystal phases identified as $\beta$-$In_2Se_3$, $\gamma$-$In_2Se_3$ and $\gamma$-InSe are determined. Moreover, it is found that the predominant impact on the crystal phase has In/Se flux ratio. The growth temperature on the other hand, at least in the 300-450 ºC range, impacts mostly the surface roughness and not the crystal phase of indium selenide. $\beta$-$In_2Se_3$ grows at In/Se flux ratio of 0.005-0.008, $\gamma$-$In_2Se_3$ – at 0.2-0.3 and $\gamma$-InSe - at 0.5-0.7. In the case of an intermediate flux ratio one obtains a mixture of different indium selenide crystalline phases. The increase of In/Se flux ratio above 0.6 leads to the precipitation of indium nano-droplets on the surface followed by the growth of indium selenide nanowires, as observed in previous reports[23].

The identification of the indium selenide crystal structure is performed based on the combined study including the Raman scattering, the low-temperature photoluminescence (PL) and the X-ray diffraction. First of all, it is found that the Raman scattering spectra from different indium selenide thin layers vary significantly depending on the In/Se flux ratio used for their growth, Figure 1b. As the position of the Raman modes is characteristic for every indium selenide crystalline phase these measurements have provided us with initial information allowing to establish the indium selenide crystal phase diagram, Figure 1a. In the case In/Se flux ratio of 0.5 three peaks are observed at 117 cm$^{-1}$, 179 cm$^{-1}$ and 228 cm$^{-1}$. They have been previously identified as two $A_1$ (117 cm$^{-1}$, 228 cm$^{-1}$) and $E_g$ (179 cm$^{-1}$) Raman modes from $\gamma$-InSe[13]. When changing the In/Se flux ratio to 0.25 and 0.025, the Raman spectrum also changes and the three modes appear at 82 cm$^{-1}$, 152 cm$^{-1}$ and 230 cm$^{-1}$. Based on these spectral positions it can be concluded that the corresponding indium selenide crystalline phase is γ-In$_2$Se$_3$[13,24]. For the lowest In/Se flux ratio of 0.008, the Raman peaks show up at 110 cm$^{-1}$, 175 cm$^{-1}$ and 205 cm$^{-1}$ which fits well to the modes observed for β-In$_2$Se$_3$[25].

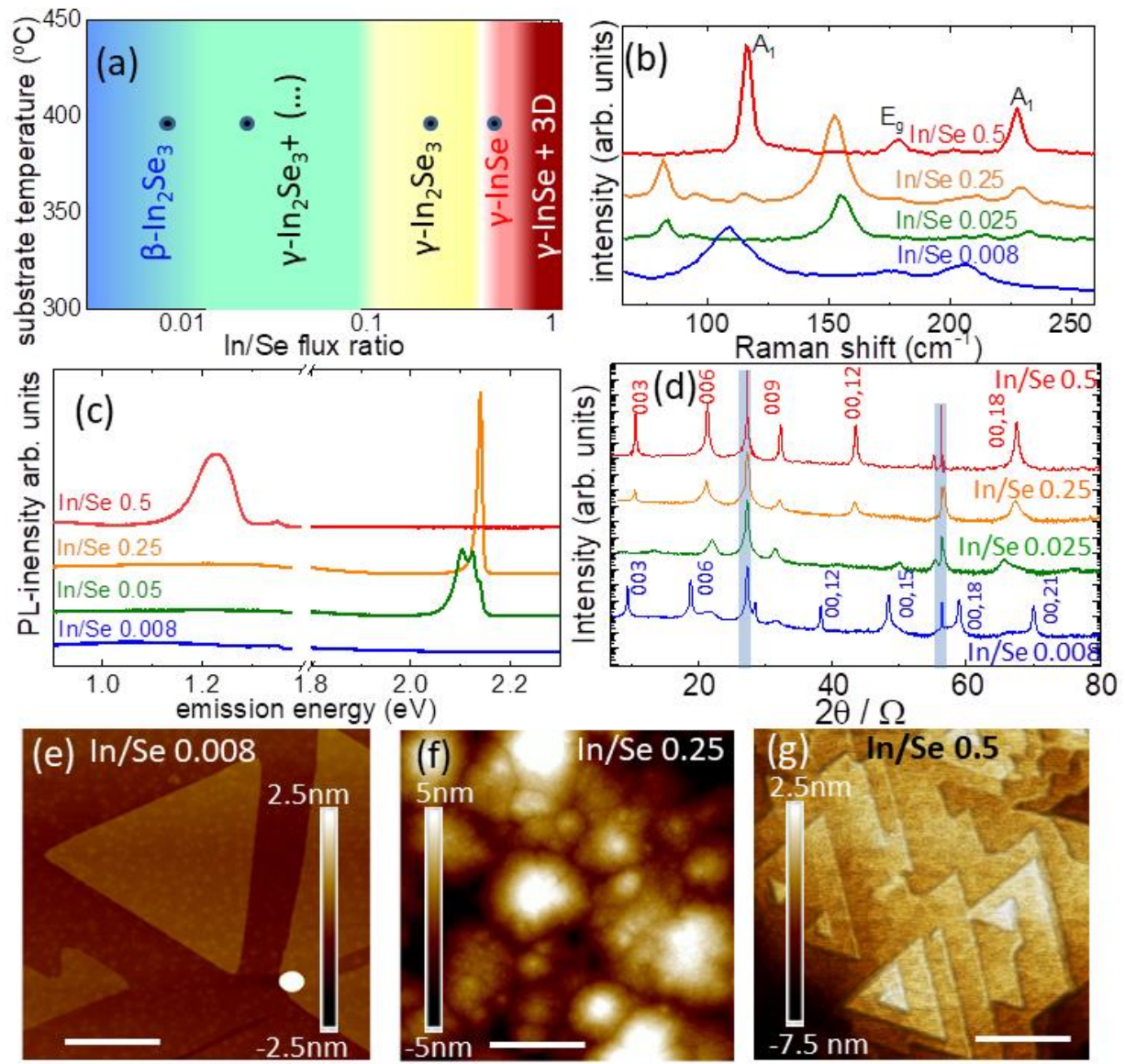

**Figure 1** (a) Indium selenide crystal phase diagram depending on MBE growth temperature and In/Se flux ratio; black points correspond to the growth conditions of the indium selenide layers for which the experimental data is shown in (b)-(d); (b) room temperature Raman scattering (c) photoluminescence measured at 10 K (the excitation line is at 405 nm) (d) X-ray diffraction: 2θ/Ω scans from the indium telluride thin layers grown at 400 ºC with different In/Se flux ratios indicated in (a). The reflexes originating from GaAs substrate are shaded in grey (e) Atomic Force Microscopy of a typical β-In$_2$Se$_3$, (f) γ-In$_2$Se$_3$ and (g) γ-InSe showing significantly different surface morphology depending on the crystal phase. Scale bars correspond to 1 μm

The identification of the crystalline phases depending on the growth conditions is confirmed by low temperature PL study, Figure 1c. For the In/Se flux ratio of 0.5, PL-spectrum from indium selenide thin layers is centered at 1.23 eV which is quite close to the reported band gap of the γ-InSe phase, which is consistent with the Raman scattering data obtained for the same sample. It has to be noted, however, that this emission appears at a slightly lower energy as compared to the band gap of bulk γ-InSe[11,13](1.3 eV). A possible interpretation of this shift relies on the strain originating from the lattice mismatch between indium selenide and GaAs substrate. In the case of γ-$In_2Se_3$ obtained for In/Se flux ratio of 0.25, the band gap is expected to change significantly and amount to 2.14 eV[13,26]. Indeed, it is found that the PL spectrum from this sample consists of only one emission line at 2.14 eV, which confirms its identification obtained on the basis of the Raman scattering. Decreasing the In/Se flux ratio to 0.025 results in the appearance of additional two emission lines in the spectral range 2.09 eV – 2.13 eV. Those lines have already been observed and are identified to be due to inclusions of other indium selenide crystalline phases within γ-$In_2Se_3$[21]. For the lowest In/Se flux ratio of 0.008, no reliable optical emission has been detected neither in the visible nor in the near infrared range. The corresponding crystal phase that has been determined by the Raman scattering is β-$In_2Se_3$. Its band gap is expected at about 1.5 eV[13,25]. In our case, however, this emission range is not accessible due to the strong PL-signal from the GaAs substrate. The intense substrate-related emission is also the reason why the corresponding spectral range is not shown in Figure 1c.

X-ray diffraction allows us for the identification of the γ-InSe and β-$In_2Se_3$, as demonstrated on the topmost and the lowest 2θ/Ω scans in Figure 1d, respectively. The γ-InSe crystal phase is confirmed again for the layers grown at In/Se flux ratio of 0.5 and the out-of-the-plane lattice constant determined from these measurements amounts to 25.95 Å, which fits well to the values reported previously[14]. It has to be noted, however, that the presence of the diffraction peaks at 2θ of 10.6º, 21.2º, 32.1º, 43.3º, as observed in the topmost diffraction pattern in Figure 1d, does not determine unambiguously the crystalline structure, since they can be assigned either to (003), (006), (008), (00,12) planes of γ-InSe or to (002), (004), (006), (008) planes of β-InSe or ε-InSe[12]. That is why the identification of the crystal phases is made by the transmission electron microscopy performed in scanning mode (STEM), Figure 2. The reflexes from the γ-$In_2Se_3$ phase characteristic for In/Se flux ratio of 0.25 are expected to overlap with that of (111)B-GaAs substrate which makes difficult to determine this phase by X-ray diffraction. On the other hand, β-$In_2Se_3$ which is characteristic for the layers grown at In/Se flux ratio of 0.008, can be well identified by the X-ray diffraction. The observed diffraction pattern differs significantly from that of γ-InSe, the lowest 2θ/Ω scan in Figure 1d. The out-of-the-plane lattice constant determined from this scan amounts to 28.20 Å which is expected based on the existing data for the β-$In_2Se_3$ phase[25,27].

Atomic force microscopy (AFM) has been used for the investigation of the surface roughness. Importantly, it reveals triangularly shaped structures on the surfaces in the case of the 2D van der Waals indium selenide crystalline phases: γ-InSe and β-$In_2Se_3$, Figure 1e and Figure 1g, respectively. On the other hand, oblique structures are typical at the surface of wurtzite-type γ-$In_2Se_3$, Figure 1f. This observation confirms independently that depending on the In/Se flux ratio indium selenide crystalline phases with significantly different symmetries can be obtained. Moreover, the changes of the growth temperature in the 300 - 450ºC range influence significantly the surface roughness without changing the indium selenide crystal phase. The typical length of the triangular structures on the surface of β-$In_2Se_3$ grown at 450 ºC amounts to 3 µm, Figure 1g, and it decreases down to 300 nm for the layers grown at 300 ºC (not shown). Importantly, when tuning properly the growth temperature one can obtain an almost flat indium selenide surface with distinct one-atomic steps with the height of about 1.0 nm, Figure 1e and 1g.

**2.2 Crystal phase heterostructures**

In the next step indium selenide crystal phase heterostructures are grown by changing the growth conditions during the MBE process in a controlled manner. Several combinations of crystal phases have been investigated. The most promising case for photonic applications is revealed by the growth of γ-InSe on γ-$In_2Se_3$ since it results in the appearance of an optical emission in the near infra-red spectral range. The MBE growth of these crystal phase heterostructures starts with 60 nm thick γ-$In_2Se_3$ on (111)B-oriented GaAs substrate. The proper growth conditions determined in the previous part of this report are used, i.e., In/Se flux ratio of 0.2 and the growth temperature of 450 ºC. The process is monitored *in situ* by the Reflection of High Energy Electron Diffraction (RHEED). After deposition of γ-$In_2Se_3$ layer a characteristic RHEED patterns for this crystal phase is observed. It is manifested by the presence of additional reflexes when the electron beam is applied along the [2,-1,-1] direction, marked by arrows in Figure 2a, upper image. Next, the growth is interrupted for 30 min and In/Se flux ratio changed to 0.6 by the reduction of Se flux. After restarting the growth one observes that RHEED pattern exhibits a distinct change and the reflexes typical for γ-$In_2Se_3$ phase disappear gradually. This is the first important hint that indium selenide crystal phase has changed due to the change of the growth conditions. After the deposition of approximately 3 nm of the second crystal phase the additional reflexes in RHEED pattern disappear completely, Figure 2a lower image.

Raman scattering from the γ-InSe/γ-$In_2Se_3$ heterostructure consists of multiple modes, the topmost spectrum in Figure 2b. In order to relate them to a particular indium selenide crystalline structure the Raman spectra for γ-InSe and γ-$In_2Se_3$ are presented separately in the Figure 2b. The comparison of these spectra let us ascribe the peaks at 83 cm$^{-1}$ and 152 cm$^{-1}$ to the γ-In$_2$Se$_3$ phase while the peaks at 117 cm$^{-1}$ and 179 cm$^{-1}$ are associated to the γ-InSe phase, confirming that, indeed, the desired heterostructure is successfully fabricated.

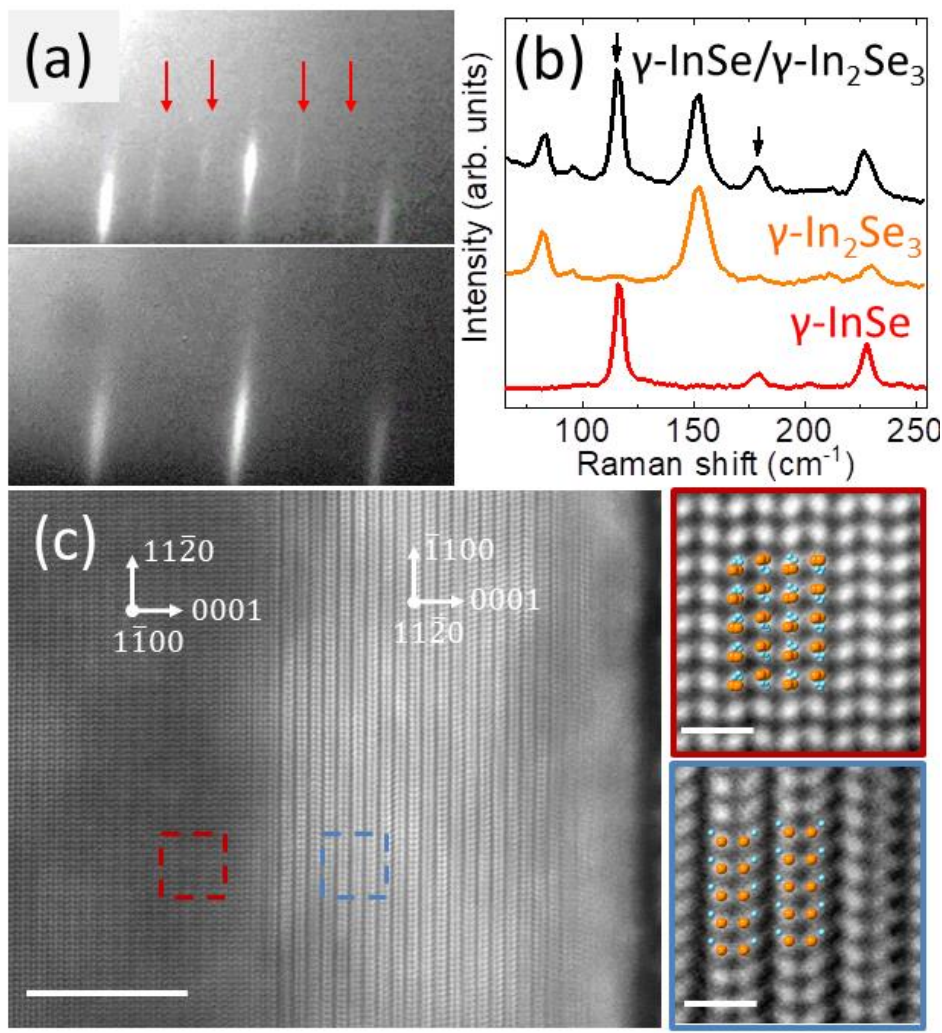

**Figure 2** Demonstration of the successful fabrication of γ-InSe/γ-In$_2$Se$_3$ indium selenide crystal phase heterostructure by various experimental techniques (a) Reflection of High Energy Electron Diffraction (RHEED) along [2,-1,-1] direction after the deposition of 60 nm γ-In$_2$Se$_3$, top panel; and after the additional 12 nm of γ-InSe, bottom panel. The presence of reflexes marked by red arrows is characteristic for the γ-In$_2$Se$_3$ phase (b) Raman scattering from γ-InSe/γ-In$_2$Se$_3$ crystal phase heterostructure (black) compared to reference γ-In$_2$Se$_3$ (orange) and γ-InSe layer (red); black arrows mark the modes related to γ-InSe on the spectrum from the heterostructure (c) Bragg-filtered STEM image evidencing the presence of two distinct indium selenide crystalline phases with a quite sharp interface in-between. In the close-ups the crystalline structures on the left (red frame) and right side (blue frame) of the image are magnified and identified as γ-In$_2$Se$_3$ and γ-InSe, respectively. Orange and blue points show the predicted positions of In and Se atoms for both indium selenide structures, respectively. Scale bar in (c) corresponds to 10 nm and the scale bars in the close-ups - to 2 nm.

The presence of two distinct indium selenide crystal lattices is evidenced by the scanning transmission electron microscopy (STEM), Figure 2c. One observes a clearly different indium selenide crystal lattice on the left side compared to the right side of Figure 2c. The left side crystalline structure is identified as wurtzite-type γ-In$_2$Se$_3$, as shown in the top close-up in Figure 2c, while on the right side the van der Waals γ-InSe phase is present, the bottom close-up in Figure 2c.

AFM study performed on γ-InSe/γ-In$_2$Se$_3$ heterostructure reveals triangularly shaped structures at the surface (not shown) which is consistent with the fact that a 2D crystal phase of indium selenide phase is the topmost layer. This finding confirms again the change of the crystal structure induced by the reduction of In/Se flux ratio during the MBE process.

Interestingly, it is found that not all possible combinations of the indium selenide crystal phases can be obtained by MBE. For instance, an attempt to grow β-In$_2$Se$_3$ phase on γ-In$_2$Se$_3$ was unsuccessful. After decrease of In/Se flux ratio from 0.2 to 0.008 the RHEED pattern does not change and exhibits typical pattern for γ-In$_2$Se$_3$. AFM and PL measurements performed on these structures do not show any indications for the presence of other phases than γ-In$_2$Se$_3$.

## 2.3 Optical emission

Importantly, it is found that the deposition of γ-InSe on γ-In$_2$Se$_3$ results in the appearance of a quite intense optical emission in the near infrared spectral region. In Figure 3a, the low temperature PL-spectrum from γ-InSe/γ-In$_2$Se$_3$ crystal phase heterostructure is compared to two references: γ-In$_2$Se$_3$ layer and γ-InSe layer grown directly on GaAs (111)B substrate. In the visible spectral range the emission spectra of γ-In$_2$Se$_3$ and γ-InSe/γ-In$_2$Se$_3$ heterostructure consist of only one emission line at 2.14 eV indicating high quality γ-In$_2$Se$_3$ crystal phase, Figure 3a. In the near infra-red, however, an additional emission line with the maximum at 1.1 eV appears in the case of the γ-InSe/γ-In$_2$Se$_3$ crystal phase heterostructure. Its energy is distinctly lower as compared to the γ-InSe related emission shown in Figure 3a in the lowest spectrum. One possible explanation of this spectral shift is that this emission could be associated to deep level defects within γ-InSe. Another explanation relies on the possibility that this emission originates from the optical transition at the γ-InSe/γ-In$_2$Se$_3$ interface. According to the band diagram shown in ref.[13], it is expected that this junction would be characterized by a staggered, type II, band alignment, as shown schematically in the inset of Figure 3c. In that case the interfacial emission energy should be lower than that of both band gaps of crystalline phases constituting the junction. The observation of the PL-line at 1.1 eV is, therefore, consistent with this interpretation.

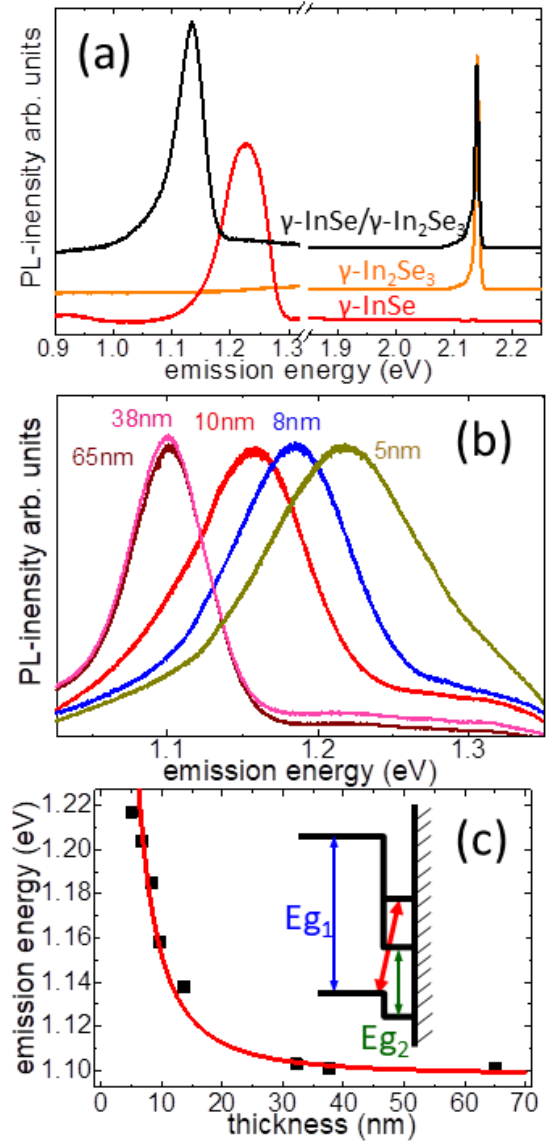

**Figure 3** Optical emission from indium selenide crystal phase heterostructures (a) Photoluminescence spectrum from γ-InSe/γ-In$_2$Se$_3$ heterostructure with the top layer thickness of 14 nm, black curve, compared to the reference γ-In$_2$Se$_3$ layer, orange curve , and γ-InSe layer, red curve; (b) Spectral shift of the emission energy induced by the quantum confinement within the γ-InSe layer assessed by photoluminescence spectra (normalized) from five crystal phase heterostructures with different thicknesses of the γ-InSe layer ranging from 65 nm to 5 nm. The temperature of the measurement is 10 K and the excitation is performed with 405 nm laser line with the fluence of 0.5 mW (c) Spectral position as a function of the layer thickness (bold squares). Solid lines represent the calculations of the quantum size effect induced spectral shift of the optical emission assuming the effective electron mass of 0.09 m$_0$ (bold line), where m$_0$ is the electron mass. The inset represents the schematic band diagram used for the calculations.

Interestingly, when changing the thickness of the γ-InSe layer in the γ-InSe/γ-In$_2$Se$_3$ heterostructure the optical emission exhibits a distinct blueshift of about 250 meV, Figure 3b. The layer thickness is determined from a comprehensive study involving scanning electron microscopy of the structure cross-section for layers thicker than 10 nm and the extrapolation of this quantity for the thinnest layers based on the growth time. Figure 3c represents the dependence of the optical emission energy as a function of γ-InSe layer thickness. When the thickness is smaller than 20 nm the aforementioned blue shift takes place. This feature is characteristic for the impact of the quantum size effect in typical medium band-gap semiconductors. Changing the thickness in the range 30 nm – 65 nm does not affect the spectral position and the shape of the optical emission which is consistent with this interpretation. The decrease of the layer thickness below 5 nm results in a significant reduction of the photoluminescence signal, so that it becomes too weak to be distinguished from the background counts. This observation is consistent with previous reports dedicated to exfoliated indium selenide layers and the expected transition from a direct to an indirect band gap semiconductor for the thinnest γ-InSe layers[28,29].

The energy blue shift as a function of the layer thickness can be well reproduced by the calculations of the optical transitions of a quantum system shown schematically in the inset of Figure 3c. Two semiconductors: γ-In$_2$Se$_3$ with the band gap of Eg$_1$ = 2.14eV and γ-InSe with the band gap of Eg$_2$ = 1.25eV form a junction characterized by a staggered band alignment whereas only the thickness of the second layer is varied. The valence band offset, VBO, is determined from the experimental value of the emission energy for large γ-InSe thicknesses, E$_d$ = 1.10 eV, by applying the simple formula: VBO = Eg$_2$ - E$_d$ = 0.15 eV. The quantum size effect takes place in the conduction band only and the best fit to the experimental values is obtained for the electron effective mass of m$_e$ = 0.09·m$_0$, where m$_0$ is the free electron mass. At the same time the conduction band offset, CBO, is quite large and amounts to 1.04 eV (CBO = Eg$_1$ - E$_d$). Therefore, the calculated impact of the quantum confinement on the emission energy in the investigated range of γ-InSe thicknesses gives similar values to the case of an electron in an infinite potential well. Importantly, it should be noted that the observation of the thickness dependence of the spectral position indicates that the optical emission is not caused by optical transitions within any deep level defect, since they are usually not sensitive to the variation of the band gap due to their local character.

The properties of the PL-line at about 1.10 eV from γ-InSe/γ-In$_2$Se$_3$ are further investigated by performing PL measurements as a function of the temperature and the excitation fluence, Figure 4. It is found that the increase of the temperature results in the fast decrease of the emission intensity and its complete disappearance already at 50 K, Figure 4a. At the same time the spectral position exhibits a distinct redshift, Figure 4b. At relatively high excitation fluence of 30 mW this dependence corresponds well to a typical temperature induced spectral shift observed for γ-In$_2$Se$_3$ excitonic emission, marked by the red line in Figure 4b. This observation indicates that the origin of the redshift is associated to the typical bandgap reduction due to the expansion of the crystal lattice with increasing temperature. The energy shift measured at relatively low excitation fluence of 0.5 mW is slightly larger than that measured for the high excitation fluence, Figure 4b. At the low excitation fluence, additional processes caused by the inhomogeneity within the γ-InSe layer, such as the thermally activated redistribution of carriers toward the quantum objects emitting at lower emission energies, may take place leading to an additional spectral shift similar to the case observed previously for self-assembled quantum dots[30,31]. Moreover, it is found that the PL-intensity increases linearly in a relatively wide range of excitation fluencies, from 30 to 1000 μW, which is, again, typical for the excitonic emission, Figure 4c-d. In contrast, sublinear increase would suggest free-to bound transitions or donor-acceptor recombination. When increasing the excitation fluence one observes a sizeable blueshift of the emission energy, Figure 4e. The latter effect is consistent with the association of this emission to optical transition at the type II γ-InSe/γ-In$_2$Se$_3$ interface. In that case, the blueshift would be induced by the electron-hole spatial separation at the interface and the resulting electric field that should increase with increasing excitation fluence. Therefore, we conclude that the optical emission at 1.10 eV is, most likely, associated to the interfacial transition at γ-InSe/γ-In$_2$Se$_3$ interface.

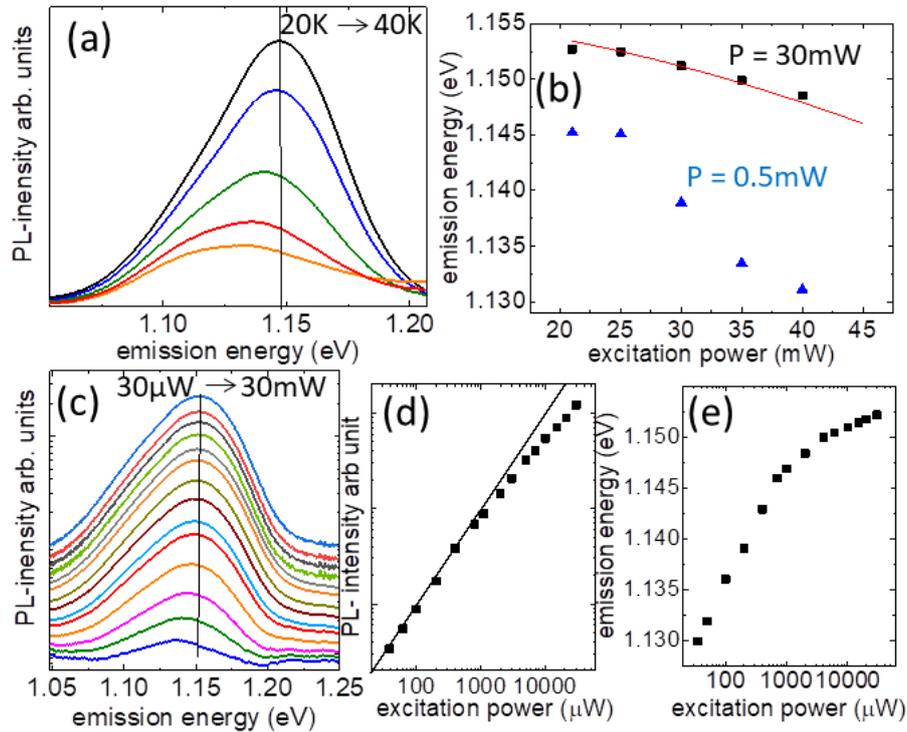

**Figure 4** (a) Temperature dependence of the optical emission from γ-InSe/γ-In$_2$Se$_3$ heterostructure with the average top layer thickness of 16 nm, excitation power 0.5 mW, (b) spectral position vs. temperature for two different excitation fluencies of 0.5mW and 30mW. The redline corresponds to thermally induced γ-In$_2$Se$_3$ bandgap shift (c) Excitation fluence dependence of the optical emission measured at 10 K (d) Increase of the photoluminescence intensity with increasing excitation fluence. Black line corresponds to the linear increase (e) Sizable blueshift of the spectral position with increasing excitation fluence; excitation is performed with 405 nm laser line.

In order to investigate inhomogeneities of the optical emission from the γ-InSe/γ-In$_2$Se$_3$ heterostructure, the laser excitation spot is reduced down to 3 µm using an optical objective in the system for micro-PL. It is found that the spectral width of the emission line does not change significantly compared to the macro-PL measurements and amounts typically to 30 meV (not shown). Depending on the spatial position the emission energy varies only slightly, by about 5 meV. Importantly, also in the case of the local excitation in the micro-PL setup, the PL-lines shift gradually towards higher energy as result of the increasing excitation fluence similar to the case shown in Figure 4c. Therefore, we conclude that this shift does not originate from local fluctuations within the γ-InSe layer which would represent an alternative explanation of this effect. Increasing the excitation fluence could populate the states with higher energies leading to the overall blueshift of the emission band. The absence of any additional lines appearing in micro-PL at high excitation fluence indicates that this possibility is not likely.

## 3. Conclusions

In summary, MBE is used to obtain the control over indium selenide crystal phase with the aim to fabricate optically active indium selenide based crystal phase heterostructures. It is found that the most important growth parameter determining the indium selenide crystal phase is the In/Se flux ratio. Varying the growth temperature in the range 350 - 400 °C is responsible mostly for the change of the surface roughness. The proper conditions for the growth of γ-InSe, γ-In$_2$Se$_3$ and β-In$_2$Se$_3$ are found. This knowledge is used to develop indium selenide crystal phase heterostructures characterized by sharp interfaces in a well-controlled manner. In particular, it is found that the γ-InSe/γ-In$_2$Se$_3$ heterostructure exhibits an optical emission in the near infrared spectral range. When decreasing the γ-InSe layer thickness below 20 nm, one observes a significant blue shift of the emission energy, even by 250 meV as compared to thick layers. It is attributed to the quantum size effect. Moreover, the optical emission from the γ-InSe/γ-In$_2$Se$_3$ crystal phase heterostructure for the largest γ-InSe thicknesses, in the range 30 nm - 65 nm, appears at distinctly lower energy as compared to the γ-InSe layer grown directly on GaAs substrate. This effect is explained by the staggered band alignment at the γ-InSe/γ-In$_2$Se$_3$ junction. In that case, our results allow for the determination of the γ-InSe/γ-In$_2$Se$_3$ valence band offset which amounts to 0.15 eV. Finally it should be noted that the well-controlled optically active crystal phase heterostructures made entirely of indium selenide having the possibility of emission energy tuning in a broad range may be used for the design of light sources and detectors in the near infra-red and visible spectral range. Particularly interesting in view of possible applications is the fact that MBE allows for the growth of thin layers on large surfaces with clean interfaces between them ensuring the repeatability of the physical properties of the investigated heterostructures.

## 4. Experimental Methods

MBE details: the samples are grown in a system for molecular beam epitaxy by PREVAC equipped with In and Se Knudsen effusion cells. The growth is monitored *in situ* using RHEED (Staib) for which the electron acceleration voltage is set to 15 kV and the electron current - to 1.5 A.

Room temperature Raman scattering measurements are performed using a Renishaw inVia Raman Microscope and a 532 nm line from Nd:YAG laser. The laser beam is focused on the sample by an x100 objective with the numerical aperture of NA=0.9 in the backscattering geometry. The laser spot size is about 0.5 μm and the average excitation power is kept below 0.1 mW to prevent heating of the sample.

For low temperature photoluminescence (PL), 'as grown' samples are placed in a closed cycle cryostat at 10 K. 405 nm laser excitation line is focused down to 0.1 mm diameter spot. To detect the signal 303 mm monochromator, SR 303i by Andor, equipped with a CCD camera, iDus CCD 401, for the visible spectral range and InGaAs detector array, iDus InGaAs 1.7, for the infrared is used. For micro-photoluminescence (μ-PL) measurements NWs are excited by 405 nm laser beam focused onto a ~3 μm diameter spot by a microscope objective. The emitted light is collected by the same objective and detected by a 500 mm-monochromator (SR-500i by Andor) equipped with a CCD camera, iDus CCD 401.

The crystalline structures of indium selenide layers were investigated using FEI Titan Cubed 80−300 transmission electron microscope operating at 300 kV. The electron transparent cross sections of the investigated samples were prepared using Focused Ion Beam (FIB) lift – off procedure in Helios Nanolab Dualbeam 600 with Omniprobe nanomanipulator. Before TEM investigation, samples were placed in plasma cleaner for 1 minute to prevent contamination. The scanning transmission electron microscopy (STEM) images were acquired for different camera length of 37mm to 230mm and collection angles ranging from 42 to 144 mrad with a converged semi-angle of 9.5 mrad of the incident beam.

## Acknowledgements

This work has been partially supported by the National Centre of Science (Poland) through grant 2021/41/B/ST3/03651, and by the Foundation for Polish Science project "MagTop" no. FENG.02.01-IP.05-0028/23 co-financed by the European Union from the funds of Priority 2 of the European Funds for a Smart Economy Program 2021–2027 (FENG). We thank Zuzanna Snioch for the technical support on the acquisition of the optical data.

**Conflict of Interest**

The authors declare no conflict of interest.

**Data Availability Statement**

The data that support the findings of this study are available from the corresponding author upon reasonable request.